\documentclass[11pt]{article}

\usepackage[a4paper,margin=1in]{geometry}
\usepackage{amsmath,amssymb,bm}
\usepackage{physics}
\usepackage{mathtools}
\usepackage{booktabs}
\usepackage{array}
\usepackage{graphicx}
\usepackage{xcolor}
\usepackage{microtype}
\usepackage{setspace}
\usepackage{cite}
\usepackage{hyperref}

\hypersetup{
    colorlinks=true,
    linkcolor=blue,
    citecolor=blue,
    urlcolor=blue
}

\numberwithin{equation}{section}

\newcommand{\ct}{\widetilde c}
\newcommand{\avg}[1]{\left\langle #1\right\rangle_{\Delta}}
\newcommand{\br}{\mathbf r}
\newcommand{\bG}{\mathbf G}

\title{\textbf{From Freezing to Terminal Packing: A Puzzle and a Paradox}}

\author{
Biman Bagchi\\
Solid State and Structural Chemistry Unit\\
Indian Institute of Science\\
Bengaluru 560012, India\\
Email: \href{mailto:profbiman@gmail.com}{\texttt{profbiman@gmail.com}}\\
}

\date{}

\begin{document}

\maketitle

\begin{abstract}

A density-functional theory developed for equilibrium freezing also produces a high-density endpoint close to terminal packing. We revisit the nonlinear Ramakrishnan--Yussouff bifurcation framework and distinguish the first crystalline fold, thermodynamic coexistence, and terminal marginality. At the terminal point, the grand-potential curvature vanishes along the principal finite-wavevector density mode, exactly when the nonlinear self-consistency equation loses stiffness. The construction gives exact close packing for one-dimensional hard rods, a terminal hard-disk packing fraction near 0.8407, and a hard-sphere value near 0.629, close to random close packing. These results identify a finite-wavevector thermodynamic marginality of a metastable liquid continuation, rather than equilibrium freezing or mechanical jamming, and expose a persistent puzzle: why should liquid-state correlations encode the approach to terminal disordered packing across different physical dimensions and systems?

\end{abstract}

\section{Introduction}
\label{sec:introduction}

Bernal's geometrical picture of liquids established the modern connection
between dense-liquid structure and irregular packing \cite{Bernal1959,Bernal1960,Bernal1964}.
For equal hard spheres, the contrast is immediate: the maximum fcc or hcp
crystalline packing fraction is $\pi/(3\sqrt{2})\simeq0.74048$
\cite{Ziman1979}, whereas mechanically generated disordered packings repeatedly
approach a limiting fraction near $0.64$. The latter value became known as the
random-close-packing density, but unlike crystalline close packing it is not
fixed by the geometry of a periodic lattice.

The modern interpretation of random close packing is necessarily
protocol-aware. Jammed densities depend on the preparation procedure, the
allowed crystalline order, friction, and the precise definition of rigidity;
the maximally random jammed formulation was introduced to separate mechanical
jamming from an ill-defined optimization of ``randomness''
\cite{TorquatoStillinger2010}. Mechanical marginality supplies a complementary
criterion: Maxwell counting gives the isostatic contact number $Z_c=2d$ for a
frictionless packing in $d$ dimensions, apart from finite-size corrections and
rattlers \cite{Maxwell1864,LiuNagel1998}. This rigidity condition does not,
however, determine the packing fraction.

Liquid-state theory approaches the problem from a different direction. A
uniform fluid is described through the radial distribution function, structure
factor, and direct correlation function related by the Ornstein--Zernike
equation \cite{HansenMcDonald1986}. These quantities accurately encode the
growth of excluded-volume correlations as a hard-sphere liquid is compressed,
but no ordinary equilibrium singularity selects the random-close-packing
density. A recent comprehensive review emphasizes both the remarkable success
of the hard-sphere model and the unresolved relation among crystallization,
glassy arrest, and jamming \cite{RoyallEtAl2024}.

\subsection{An unexpected question for a theory of freezing}

Classical density-functional theory connects the freezing transition to the spontaneous appearance of periodic components in the one-particle density. In the Ramakrishnan--Yussouff (RY) formulation, the grand potential is expanded about a uniform reference liquid, and the liquid direct correlation function provides the structural kernel that controls the response to density modulation \cite{RamakrishnanYussouff1979}. The order parameters are the Fourier coefficients of the inhomogeneous density at reciprocal-lattice wavevectors. Oxtoby's Les Houches review placed this liquid-based density-functional description in the broader theory of crystallization, including its treatment of interfaces and nucleation \cite{Oxtoby1991}. The framework was constructed to describe equilibrium crystallization and its thermodynamics; it contains no explicit contact network, rigidity criterion, or compression protocol for producing a jammed state.

The subsequent development of density-wave freezing theory established both its reach and its limitations. Ramakrishnan showed that a finite-amplitude hexagonal density wave can describe a strongly first-order freezing transition in two dimensions \cite{Ramakrishnan1982PRL}. Bagchi, Cerjan, and Rice (BCR) reformulated the nonlinear self-consistency problem as a bifurcation analysis and applied it to hard spheres \cite{BagchiCerjanRice1983}. Cerjan and Bagchi later compared fcc, bcc, and simple-cubic branches and connected their local stability to the slopes of the nonlinear solution curves \cite{CerjanBagchi1985}. Fundamental-measure theory subsequently achieved dimensional crossover together with an accurate description of hard-sphere freezing and the fcc solid \cite{RosenfeldEtAl1997}. Related density-functional approaches were also extended from periodic crystals to amorphous states. Singh, Stoessel, and Wolynes examined the stability of an aperiodic density wave built on Bernal-like random packing \cite{SinghStoesselWolynes1985}, while Kim and Munakata found a metastable amorphous free-energy minimum in a hard-sphere density-functional landscape \cite{KimMunakata2003}. These works show that a liquid-based functional can encode dense ordered and amorphous structures, but they do not make it obvious why a crystalline bifurcation equation should locate terminal packing.

Two subsequent extensions of the BCR program are central to the present
reassessment.  A nonlinear diffusion theory showed that the kinetic stability
of a supercooled liquid is governed by coupled relaxation rates of several
reciprocal-lattice density modes and revealed markedly different dynamical
constraints on fcc and bcc ordering \cite{Bagchi1987PhysLett,Bagchi1987Physica}.
That analysis also found a high-density solution with vanishing crystalline
amplitudes, a finite uniform density shift, and a simultaneous zero of all the
generalized relaxation rates.  For hard spheres it occurred near
$\rho_\ell\sigma^3\simeq1.20$ and was set aside because it lay outside the
physically accessible equilibrium-fluid regime.  Its proximity to the
random-close-packing density motivates a different reading: the inaccessibility
may be a signature of a terminal metastable continuation rather than evidence
that the mathematical endpoint is irrelevant.

Independently, Bagchi and Rice extended the bifurcation analysis to hard
hyperspheres in $d$ dimensions \cite{BagchiRice1988}.  They showed that the
two-body density-functional representation is exact both for $d=1$ and in the
limit $d\rightarrow\infty$, and argued that the continuous simple-hypercubic
bifurcation at $\lambda_G=1$ identifies maximum achievable packing in both
limits.  This result gives dimensional continuation a more fundamental role
than a numerical comparison among one, two, and three dimensions: the two
exact limits constrain any interpretation of the intermediate-dimensional
RCP-like densities. Complementary calculations of hard hyperspheres found
first-order freezing to pre-empt the Kirkwood instability across the dimensions
examined and developed a fundamental-measure functional for the same class of
systems \cite{FinkenSchmidtLowen2002}.

\begin{samepage}
Throughout this work, $\rho_\ell$ and $\eta_\ell$ denote the number density and packing fraction of the uniform reference liquid, while $\rho_s$ and $\eta_s$ denote the corresponding mean quantities on the inhomogeneous branch. The liquid direct correlation function in Fourier space is $\widetilde c(k;\rho_\ell)$, and we use the shorthand $c_G\equiv\widetilde c(G;\rho_\ell)$ at the principal reciprocal-lattice wavevector. The coupling is $\lambda_G\equiv\rho_s c_G$. A superscript $T$ denotes evaluation at the terminal point; thus $c_G^{T}=\widetilde c(G;\rho_\ell^{T})$ and $\lambda_G^{T}\equiv\rho_s^{T}c_G^{T}$.

The surprising BCR observation is that continuation beyond ordinary freezing produces the terminal condition
\begin{equation}
\lambda_G^{T}\equiv\rho_s^{T}c_G^{T}=1,
\label{eq:intro_marginal_condition}
\end{equation}
\end{samepage}
For three-dimensional fcc hard spheres, a nonzero branch returns to the uniform solution at this value, and the associated terminal liquid packing fraction is approximately
\begin{equation}
\eta_\ell^{T}\simeq0.629,
\label{eq:intro_3d_density}
\end{equation}
close to the conventional random-close-packing regime. The accompanying inhomogeneous-branch density lies much nearer crystalline close packing. Thus a theory driven by equilibrium liquid correlations produces two density scales: one suggestive of terminal disordered packing and another associated with the crystal.

The lower-density fold must not be identified with equilibrium freezing.  It
is the point at which finite-amplitude crystalline stationary solutions first
appear, analogous to a turning point of a van der Waals loop.  In the one-mode
fcc theory the fold lies at $\lambda_G=0.972991$, where the crystalline grand
potential is still higher than the liquid value by only
$0.004328\,Nk_{\mathrm B}T$.  The projected Maxwell crossing occurs nearby at
$\lambda_G=0.979014$.  BCR therefore obtained a remarkably accurate estimate
of freezing from the bifurcation point, but the bifurcation and thermodynamic
coexistence conditions are conceptually distinct.

This numerical proximity would be easy to dismiss if it occurred only in one approximate three-dimensional calculation. The dimensional evidence makes that dismissal less satisfactory.

\subsection{The dimensional evidence}

In one dimension, hard rods do not undergo an ordinary liquid--solid transition. Nevertheless, the corresponding nonlinear equation has a single bifurcation at
\begin{equation}
\rho_\ell^{T}\sigma=\rho_s^{T}\sigma=1,
\label{eq:intro_1d_exact}
\end{equation}
which is the exact geometrical limit at which rods fill the line. Here the association between $\lambda_G^{T}=1$ and terminal packing is exact, not numerical. The signed solutions $\psi_G$ and $-\psi_G$ in a cosine representation are related by a translation through half a wavelength; the physical scalar measure of order is therefore $|\psi_G|$. The negative part of the algebraic diagram is not a distinct thermodynamic phase.

Two dimensions provide two complementary results that must be kept logically separate. First, Radloff \textit{et al.} solved the two-order-parameter equations for the hexagonal representation of the classical two-dimensional one-component plasma \cite{Radloff1984}. Their solution curves pass through
\begin{equation}
(\lambda_{G_\alpha},\psi_{G_\alpha})=(1,0)
\label{eq:intro_2d_ocp_point}
\end{equation}
for several values of the coupling to the second reciprocal-lattice star. The global topology changes with that coupling, and the point at unity is not an fcc-like returning bifurcation. This calculation establishes the persistence of the marginal solution under multimode coupling, not the universality of a second bifurcation.

Second, the density mapping relevant to disordered packing comes from hard disks rather than from the OCP diagram. Baus and Colot constructed an accurate rescaled representation of the direct correlation function for hard rods, disks, spheres, and hyperspheres \cite{BausColot1987}. Using this type of hard-disk structural input, Xu and Rice located the deeply metastable hard-disk stability limit near \cite{XuRice2010}
\begin{equation}
\eta_\ell^{T}\simeq0.8407,
\qquad
\phi_0\equiv\frac{\rho_s-\rho_\ell}{\rho_\ell}\simeq0.003,
\qquad
S(G)\simeq300.
\label{eq:intro_2d_density}
\end{equation}
The density discontinuity is exceptionally small and the structural peak exceptionally large, so that $\rho_s\widetilde c(G)\simeq1$. The resulting liquid packing fraction lies close to the commonly identified maximally random jammed regime for monodisperse hard disks. The agreement is suggestive, although the numerical value of a jammed density remains sensitive to protocol and to the chosen criterion of randomness.

Three dimensions add a further distinction between the marginal solution and bifurcation topology. The fcc equations possess the returning nonzero branch associated with the RCP-like density. The two-order-parameter bcc equations used in the analysis of sodium freezing also pass through $(\lambda_G,\psi_G)=(1,0)$, but the diagram is asymmetric and possesses only the lower-density freezing bifurcation \cite{CerjanBagchi1985}. Thus neither the existence of the uniform solution nor the vanishing of its linear stiffness guarantees a second bifurcation. The latter depends on lattice symmetry, relative phases, mode coupling, and the nonlinear invariants admitted by the reciprocal-lattice representation.

The dimensional evidence can therefore be summarized as follows. In one dimension, the marginal condition coincides exactly with terminal packing and with the only bifurcation. In two-dimensional hard disks, it maps to a density close to the maximally random jammed regime, while the Radloff OCP calculation shows that the solution at unity need not be a returning bifurcation. In three-dimensional fcc hard spheres, it accompanies a returning branch and selects an RCP-like liquid density, whereas bcc provides a counterexample to the same topology. What appears robust is the finite-wavevector marginal condition; what is not robust is its nonlinear realization as a second bifurcation.

\subsection{The thermodynamic puzzle and the scope of this work}

Equation~\eqref{eq:intro_marginal_condition} is not the ordinary instability condition of the uniform liquid. From the Ornstein--Zernike relation,
\begin{equation}
S(G)=\frac{1}{1-\rho_\ell\widetilde c(G)},
\label{eq:intro_structure_factor}
\end{equation}
a conventional finite-wavevector spinodal would require $\rho_\ell\widetilde c(G)=1$ and hence a divergence of $S(G)$. The BCR condition instead contains $\rho_s$. A finite liquid--solid density discontinuity can therefore complete the factor required for $\rho_s\widetilde c(G)=1$ while $S(G)$ remains finite. The distinction is small in the two-dimensional hard-disk calculation but appreciable in the three-dimensional hard-sphere problem.

The first objective of this work is to separate the lower-density fold from
thermodynamic coexistence and from the terminal point at $\lambda_G^{T}=1$.  On
continuation away from the fold, it generates a metastable crystalline
minimum and a lower barrier branch;
coexistence requires equal chemical potential and pressure; and the terminal
point is a local marginality of the returning branch.  We then identify the
thermodynamic quantity that becomes marginal at the terminal point.  When the
uniform density shift is retained consistently, the curvature of the RY grand
potential with respect to a finite-$G$ density modulation is
\begin{equation}
K_G^{\mathrm{RY}}
=
\frac{\rho_\ell^2D_G}{\rho_s}(1-\lambda_G),
\label{eq:intro_curvature_result}
\end{equation}
where $D_G>0$ is the normalization of the lattice function. The curvature therefore vanishes precisely at Eq.~\eqref{eq:intro_marginal_condition}. The local branch-stability factor obtained from the nonlinear bifurcation equation has the same sign, placing the Cerjan--Bagchi stability analysis directly inside the RY thermodynamic framework.

The second objective is to connect this static result with the earlier nonlinear
dynamics while separating both from the deeper packing question.  The
density-shifted condition that nulls the RY curvature is also the condition at
which the coupled generalized relaxation rates vanish in the mean-field
nonlinear diffusion theory.  This establishes a static and dynamical
density-wave marginality, but it does not prove Maxwell isostaticity,
amorphous localization, or a unique random-close-packed state.  We therefore
ask a narrower but sharper question: why does the same condition give exact
terminal packing in one dimension, constrain closest packing as
$d\rightarrow\infty$, and select densities close to maximally random packing
in two and three dimensions?

The remainder of the paper first summarizes the integral-equation and RY
framework and reconstructs the nonlinear bifurcation equations.  The one-,
two-, and three-dimensional solution structures are then compared, including
the thermodynamic organization of the one-mode fcc branches.  We next derive
the RY curvature and establish its equivalence to the local branch-stability
criterion.  The dimensional and dynamical results are then reassessed before
contact and finite-wavevector structure are examined as possible links to
terminal packing.

\section{Integral-Equation Theory and the Ramakrishnan--Yussouff Functional}

\subsection{Ornstein--Zernike relation and finite-wavevector structure}

For a homogeneous isotropic liquid of number density $\rho_\ell$, the total correlation function
\begin{equation}
h(r)=g(r)-1
\label{eq:h_def}
\end{equation}
is related to the direct correlation function $c(r)$ by the Ornstein--Zernike equation,
\begin{equation}
h(\br)
=
c(\br)
+
\rho_\ell
\int d\br'\,
c(|\br-\br'|)h(\br').
\label{eq:oz_real}
\end{equation}
With the convention $\widetilde f(\mathbf{k})=\int d\br\,e^{-i\mathbf{k}\cdot\br}f(\br)$, Fourier transformation gives
\begin{equation}
\widetilde h(k)
=
\frac{\ct(k)}
{1-\rho_\ell\ct(k)}.
\label{eq:oz_fourier}
\end{equation}
The static structure factor is
\begin{equation}
S(k)
=
1+\rho_\ell\widetilde h(k)
=
\frac{1}{1-\rho_\ell\ct(k)}.
\label{eq:sk}
\end{equation}

Equation~\eqref{eq:sk} distinguishes the zero-wavevector and finite-wavevector sectors. At $k=0$,
\begin{equation}
S(0)=\rho_\ell k_{\mathrm B}T\kappa_T,
\label{eq:s0_again}
\end{equation}
and a divergence signals an ordinary compressibility instability. At a nonzero wavevector $G$, a condition
\begin{equation}
1-\rho_\ell\ct(G)=0
\label{eq:finite_g_div}
\end{equation}
would imply a divergence of $S(G)$ and a linear instability of the uniform liquid to a density wave of wavelength $2\pi/G$.

This finite-$G$ criterion provides an attractive linear-instability picture of freezing \cite{RamakrishnanYussouff1979}. A dense liquid develops a pronounced first peak in $S(k)$, reflecting the dominant nearest-neighbor packing wavelength, and one might imagine freezing when that peak diverges. For hard spheres, however, the principal peak remains finite: freezing is first order and occurs before the uniform liquid reaches such an instability. Thus the condition
\begin{equation}
\rho_\ell\ct(G)=1
\label{eq:munakata_like}
\end{equation}
is not realized as an ordinary liquid spinodal in the physical density range.

The BCR condition is different:
\begin{equation}
\rho_s\ct(G)=1.
\label{eq:bcr_different}
\end{equation}
The distinction between Eqs.~\eqref{eq:munakata_like} and \eqref{eq:bcr_different} is not a minor replacement of one density by another. It is the key to the free-energy interpretation developed below.

\subsection{Ramakrishnan--Yussouff grand-potential functional}

The Ramakrishnan--Yussouff theory expands the excess free energy of an inhomogeneous state about a uniform reference liquid. Let
\begin{equation}
\delta\rho(\br)=\rho(\br)-\rho_\ell.
\label{eq:delta_rho}
\end{equation}
To second order in $\delta\rho$, the grand-potential difference between the inhomogeneous state and the reference liquid is
\begin{align}
\beta\Delta\Omega[\rho]
={}&
\int d\br\,
\left[
\rho(\br)\ln\frac{\rho(\br)}{\rho_\ell}
-\delta\rho(\br)
\right]
\nonumber\\
&-
\frac{1}{2}
\int d\br\,d\br'\,
\delta\rho(\br)
c(|\br-\br'|)
\delta\rho(\br').
\label{eq:ry_functional}
\end{align}
Here $\beta=(k_{\mathrm B}T)^{-1}$ and $c(r)$ is the direct correlation function of the reference liquid at density $\rho_\ell$.

The first line of Eq.~\eqref{eq:ry_functional} is the exact ideal-gas contribution relative to the reference liquid. The second line is the quadratic excess contribution. The molecular structure of the reference liquid enters entirely through $c(r)$. For hard spheres, all energetic scales are entropic, but the functional remains nontrivial because excluded-volume correlations are encoded in $c(r)$.

The functional in Eq.~\eqref{eq:ry_functional} is a grand-potential difference at the chemical potential of the reference liquid. Once both candidate density profiles satisfy the appropriate stationarity conditions, equilibrium coexistence requires
\begin{equation}
\Delta\Omega=0.
\label{eq:coexistence}
\end{equation}
This is the density-functional statement of equal grand potentials at common temperature and chemical potential. It must be distinguished from a curvature condition: $\Delta\Omega=0$ compares two stationary states, whereas a zero eigenvalue of the second variation identifies local marginality in a specified direction in density space.

\subsection{Uniform density shift and finite-\texorpdfstring{$G$}{G} modulation}

To connect the RY functional to the BCR construction, it is useful to separate a uniform density shift from a periodic density modulation. In a one-mode description, write
\begin{equation}
\rho(\br)
=
\rho_\ell
\left[
1+\phi_0+u_G\zeta_G(\br)
\right].
\label{eq:one_mode_density}
\end{equation}
The function $\zeta_G(\br)$ is a real, symmetry-adapted combination of reciprocal-lattice waves belonging to a star of vectors of magnitude $G$. Its unit-cell average is taken to vanish,
\begin{equation}
\avg{\zeta_G}=0,
\label{eq:zeta_mean_zero}
\end{equation}
and its normalization is defined by
\begin{equation}
D_G=\avg{\zeta_G^2}.
\label{eq:DG}
\end{equation}
No assumption that $D_G=1$ is required. In this representation, $u_G$ is the density-wave coefficient measured relative to $\rho_\ell$. It is related to the normalized BCR amplitude used in Sec.~3 by
\begin{equation}
u_G=(1+\phi_0)\psi_G=\frac{\rho_s}{\rho_\ell}\psi_G.
\label{eq:u_psi_relation}
\end{equation}
Both coefficients are signed rather than intrinsically nonnegative magnitudes of order. When a translation changes $\zeta_G$ to $-\zeta_G$, states with opposite signs describe translated density patterns. In a multimode expansion, however, relative phases and signs between different reciprocal-lattice stars can be physically significant.

The unit-cell average of Eq.~\eqref{eq:one_mode_density} is
\begin{equation}
\avg{\rho(\br)}
=
\rho_\ell(1+\phi_0).
\label{eq:average_density}
\end{equation}
Identifying this average with the density $\rho_s$ of the inhomogeneous branch gives
\begin{equation}
1+\phi_0
=
\frac{\rho_s}{\rho_\ell}.
\label{eq:density_jump}
\end{equation}
Thus $\phi_0$ represents the uniform density discontinuity between the reference liquid and the inhomogeneous branch. It becomes the liquid--solid density discontinuity only when those branches are identified with coexisting phases. Equivalently, the periodic density is $\rho_\ell u_G\zeta_G=\rho_s\psi_G\zeta_G$: the normalized BCR density wave is superposed on a branch whose mean density is $\rho_s$.

For later convenience, define
\begin{equation}
A(\br)
=
1+\phi_0+u_G\zeta_G(\br).
\label{eq:A_def}
\end{equation}
Physical density profiles require $A(\br)>0$ throughout the unit cell.
Then
\begin{equation}
\rho(\br)=\rho_\ell A(\br)
\label{eq:rho_A}
\end{equation}
and
\begin{equation}
\delta\rho(\br)
=
\rho_\ell
\left[
\phi_0+u_G\zeta_G(\br)
\right].
\label{eq:delta_rho_one_mode}
\end{equation}

The notation $A(\br)$ is particularly useful because the ideal contribution involves the nonlinear combination $A\ln A$. The curvature calculation in the following section will depend on the simple identity
\begin{equation}
\frac{\partial A(\br)}{\partial u_G}
=
\zeta_G(\br).
\label{eq:dA_dpsi}
\end{equation}

\subsection{One-mode form of the RY functional}

Substitution of Eq.~\eqref{eq:rho_A} into the ideal part of Eq.~\eqref{eq:ry_functional} gives
\begin{equation}
\frac{\beta\Delta\Omega_{\mathrm{id}}}{V}
=
\rho_\ell
\avg{
A(\br)\ln A(\br)
-
\left[A(\br)-1\right]
}.
\label{eq:ideal_A}
\end{equation}
Using Eqs.~\eqref{eq:zeta_mean_zero} and \eqref{eq:DG}, together with the orthogonality of the uniform and finite-$G$ modes, the excess contribution becomes
\begin{equation}
\frac{\beta\Delta\Omega_{\mathrm{ex}}}{V}
=
-\frac{\rho_\ell^2}{2}
\left[
\ct(0)\phi_0^2
+
D_G\ct(G)u_G^2
\right].
\label{eq:excess_one_mode}
\end{equation}
Therefore,
\begin{align}
\frac{\beta\Delta\Omega}{V}
={}&
\rho_\ell
\avg{
A(\br)\ln A(\br)
-
\left[A(\br)-1\right]
}
\nonumber\\
&-
\frac{\rho_\ell^2}{2}
\left[
\ct(0)\phi_0^2
+
D_G\ct(G)u_G^2
\right].
\label{eq:ry_one_mode}
\end{align}

Equation~\eqref{eq:ry_one_mode} contains two components of the density parametrization that must be kept conceptually distinct. The uniform component $\phi_0$ changes the mean density from $\rho_\ell$ to $\rho_s$, while $u_G$ describes a periodic density modulation relative to $\rho_\ell$. After stationarity has been imposed, coexistence follows by comparing the functional on the liquid and inhomogeneous branches; finite-$G$ marginality follows from the curvature with respect to $u_G$. The BCR equation below uses the normalized amplitude $\psi_G=u_G/(1+\phi_0)$.

The later curvature analysis will show that the factor $1/\rho_s$ arises from the ideal term in Eq.~\eqref{eq:ideal_A}. This is the origin of the BCR combination $\rho_s\ct(G)$.

\section{Bifurcation Analysis}

\subsection{Projected nonlinear equation}

The BCR construction projects the stationarity equation of the density functional onto symmetry-adapted reciprocal-lattice functions. Using the density parametrization introduced in Sec.~2, let $\alpha$ label a star of reciprocal-lattice vectors of magnitude $G_\alpha$. The corresponding structural coupling is
\begin{equation}
\lambda_\alpha
=
\rho_s\ct(G_\alpha;\rho_\ell),
\label{eq:lambda_alpha}
\end{equation}
where the argument $\rho_\ell$ emphasizes that the direct correlation function is evaluated for the uniform reference liquid, whereas $\rho_s=\rho_\ell(1+\phi_0)$ is the mean density of the inhomogeneous branch. This mixing of the two densities is essential to the BCR condition.

Projection gives coupled nonlinear equations of the form
\begin{equation}
D_\alpha\psi_\alpha
=
\frac{
\displaystyle
\int_{\Delta} d\br\,
\zeta_\alpha(\br)
\exp\left[
\sum_\beta
\lambda_\beta\psi_\beta
\zeta_\beta(\br)
\right]
}{
\displaystyle
\int_{\Delta} d\br\,
\exp\left[
\sum_\beta
\lambda_\beta\psi_\beta
\zeta_\beta(\br)
\right]
},
\label{eq:bcr_multimode}
\end{equation}
where $\Delta$ denotes a primitive unit cell and
\begin{equation}
D_\alpha
=
\avg{\zeta_\alpha^2}.
\label{eq:Dalpha}
\end{equation}
The numerical factors depend on the normalization of the lattice functions, but the nonlinear structure is unchanged.

In a one-order-parameter reduction, Eq.~\eqref{eq:bcr_multimode} becomes
\begin{equation}
D_G\psi_G
=
F_G(\lambda_G\psi_G),
\label{eq:bcr_one_mode}
\end{equation}
where
\begin{equation}
F_G(x)
=
\frac{
\avg{\zeta_G e^{x\zeta_G}}
}{
\avg{e^{x\zeta_G}}
}.
\label{eq:FG_def}
\end{equation}

Because $\avg{\zeta_G}=0$, the uniform solution $\psi_G=0$ exists for every $\lambda_G$; a nonzero solution describes a periodic density modulation.

\subsection{Linearization and the condition \texorpdfstring{$\lambda_G=1$}{lambdaG=1}}

Linearization at the uniform solution uses only $\avg{\zeta_G}=0$ and $\avg{\zeta_G^2}=D_G$:
\begin{equation}
F_G(\lambda_G\psi_G)
=
\lambda_G D_G\psi_G
+\mathcal O(\psi_G^2).
\label{eq:FG_linear}
\end{equation}
Substitution into Eq.~\eqref{eq:bcr_one_mode} gives, for $D_G\ne0$,
\begin{equation}
(1-\lambda_G)\psi_G=0.
\label{eq:marginal_linear}
\end{equation}
Thus the uniform solution is linearly marginal at
\begin{equation}
\lambda_G=1.
\label{eq:lambda_one}
\end{equation}

Equation~\eqref{eq:lambda_one} is a necessary local condition for a nonzero branch to meet the uniform branch, but it does not determine the nonlinear topology. In particular, marginality at $(\lambda_G,\psi_G)=(1,0)$ need not appear as a returning second bifurcation. Nor does the linear result identify coexistence or an ordinary liquid spinodal. Its thermodynamic meaning is established by the RY curvature calculation in Sec.~4.

\subsection{The fcc bifurcation diagram}

\begin{figure}[!h]
    \centering
    \includegraphics[width=0.82\textwidth]
    {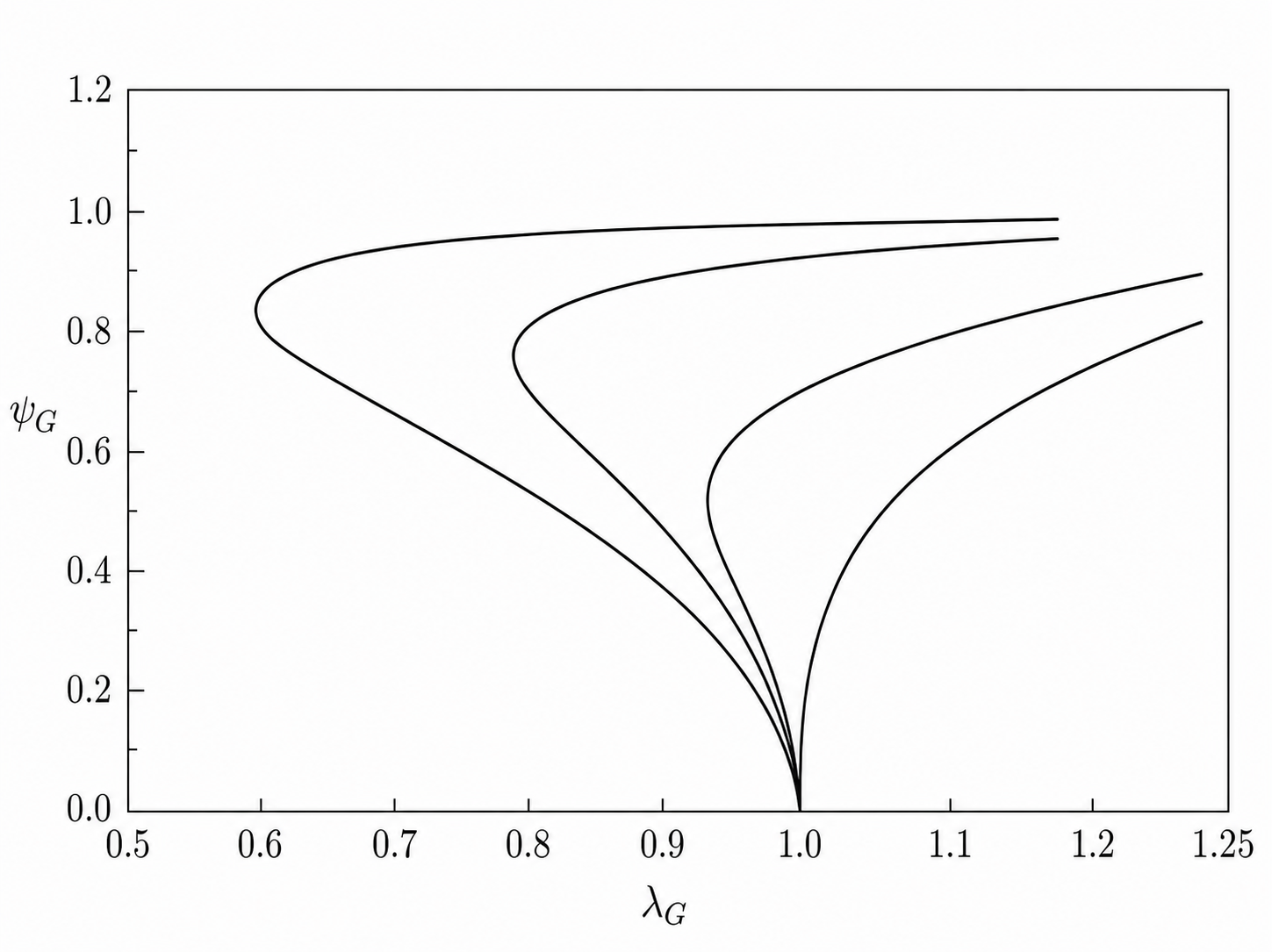}
    
\caption{Bifurcation curves obtained from the nonlinear density-functional
    equations for a three-dimensional fcc representation. The axes are the
    principal crystalline density-wave amplitude $\psi_G$ and the structural
    coupling $\lambda_G=\rho_s\widetilde{c}(G)$. Reading the upper portions of
    the four curves from left to right, the secondary-mode parameter pairs are
    $(0.25,0.10)$, $(0.15,0.08)$, $(0.05,0.05)$, and $(-2.0,-2.0)$,
    respectively. A returning branch approaches the uniform solution and ends
    at the marginal point $\lambda_G=1$; stability must be determined from the
    thermodynamic functional rather than from the plotted slope alone.
    Redrawn from the bifurcation analysis of Bagchi, Cerjan, and Rice
    \cite{BagchiCerjanRice1983}.}
    \label{fig:bifurcation_3D}
\end{figure}

For the fcc hard-sphere representation, the nonlinear BCR equations generate the two-branch structure shown in Fig.~\ref{fig:bifurcation_3D}. The lower-density bifurcation belongs to the ordinary freezing problem. More unusually, a nonzero branch returns to the uniform solution at $\lambda_G=1$.


The diagram is a mean-field solution structure, not a pressure--volume isotherm. Its vertical coordinate is a Fourier coefficient of the crystalline density. When a translation reverses the chosen lattice function, $\psi_G$ and $-\psi_G$ describe translated versions of the same density pattern, so the scalar degree of order is $|\psi_G|$. Relative signs and phases can nevertheless matter in a multimode theory. The existence of the returning branch is therefore a statement about nonlinear symmetry and mode coupling, not a generic consequence of the ever-present solution $\psi_G=0$.


\subsection{One-dimensional hard-rod bifurcation}

\begin{figure}[!h]
    \centering
    \includegraphics[width=0.78\textwidth]
    {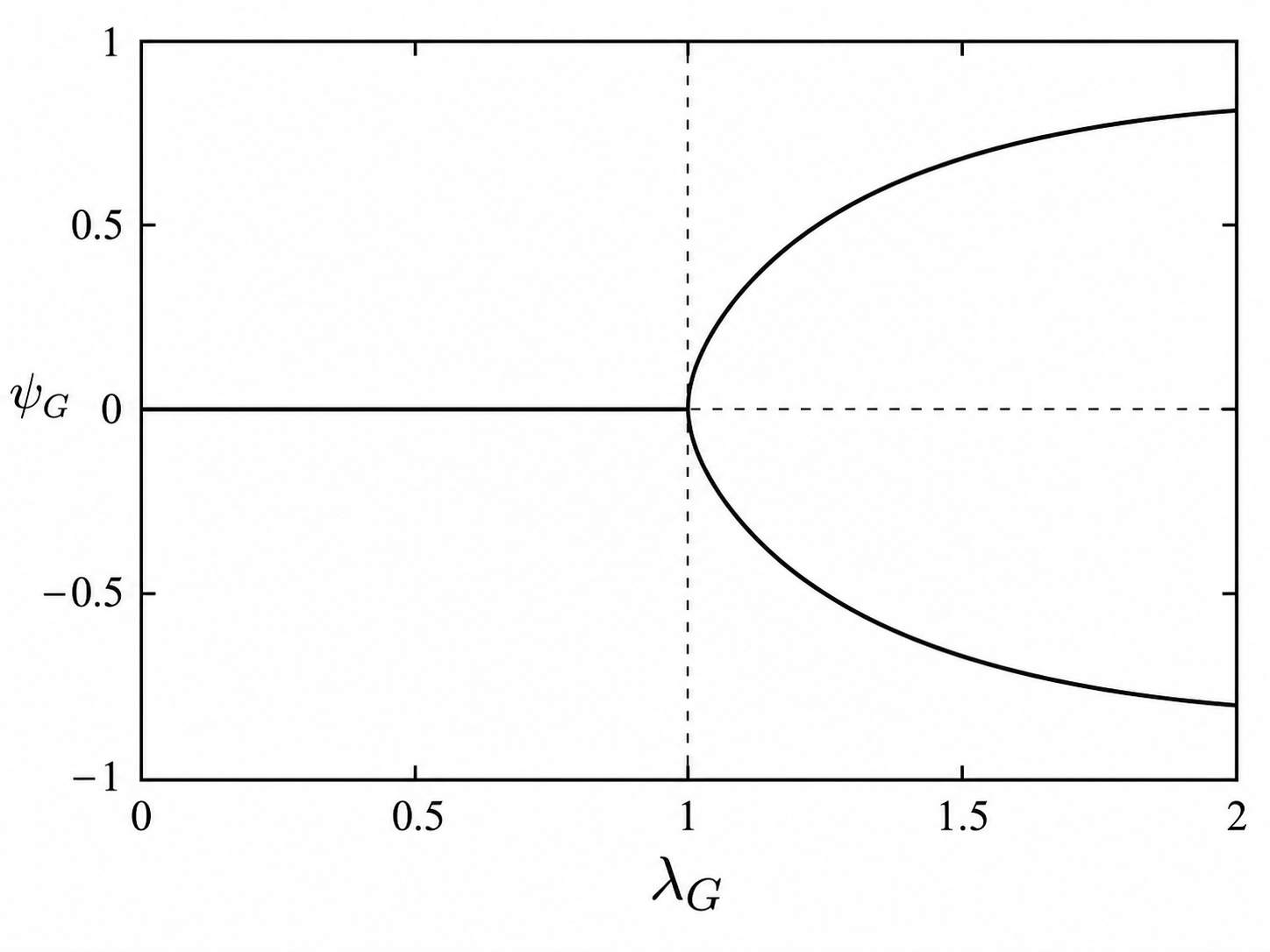}
    \caption{Bifurcation structure for one-dimensional hard rods. The
    symmetry-related positive and negative density-wave branches emerge
    continuously from the uniform solution at $\lambda_G=1$. With the exact
    one-dimensional correlation input, this marginal condition gives the exact
    terminal-packing density $\rho_\ell^{T}\sigma=\rho_s^{T}\sigma=1$
    \cite{BagchiCerjanRice1983}.}
    \label{fig:bifurcation_1D}
\end{figure}

The one-dimensional hard-rod result is the sharpest limiting case. Hard rods have no equilibrium freezing transition, yet the nonlinear equation has a single bifurcation, at $\lambda_G=1$. With the exact hard-rod correlation input, the associated densities are
\begin{equation}
\rho_\ell^{T} \sigma=\rho_s^{T}\sigma=1,
\end{equation}
the exact close-packed density \cite{BagchiCerjanRice1983}. The corresponding structure is shown in Fig.~\ref{fig:bifurcation_1D}.

The result is stronger than numerical proximity to a conventional packing fraction: the terminal density is unique, and the only bifurcation coincides exactly with exhaustion of the available packing space. It is this exact one-dimensional result that makes the near coincidence in two and three dimensions difficult to dismiss outright.

\subsection{The bcc comparison}

\begin{figure}[!htbp]
    \centering
    \includegraphics[width=0.72\textwidth]{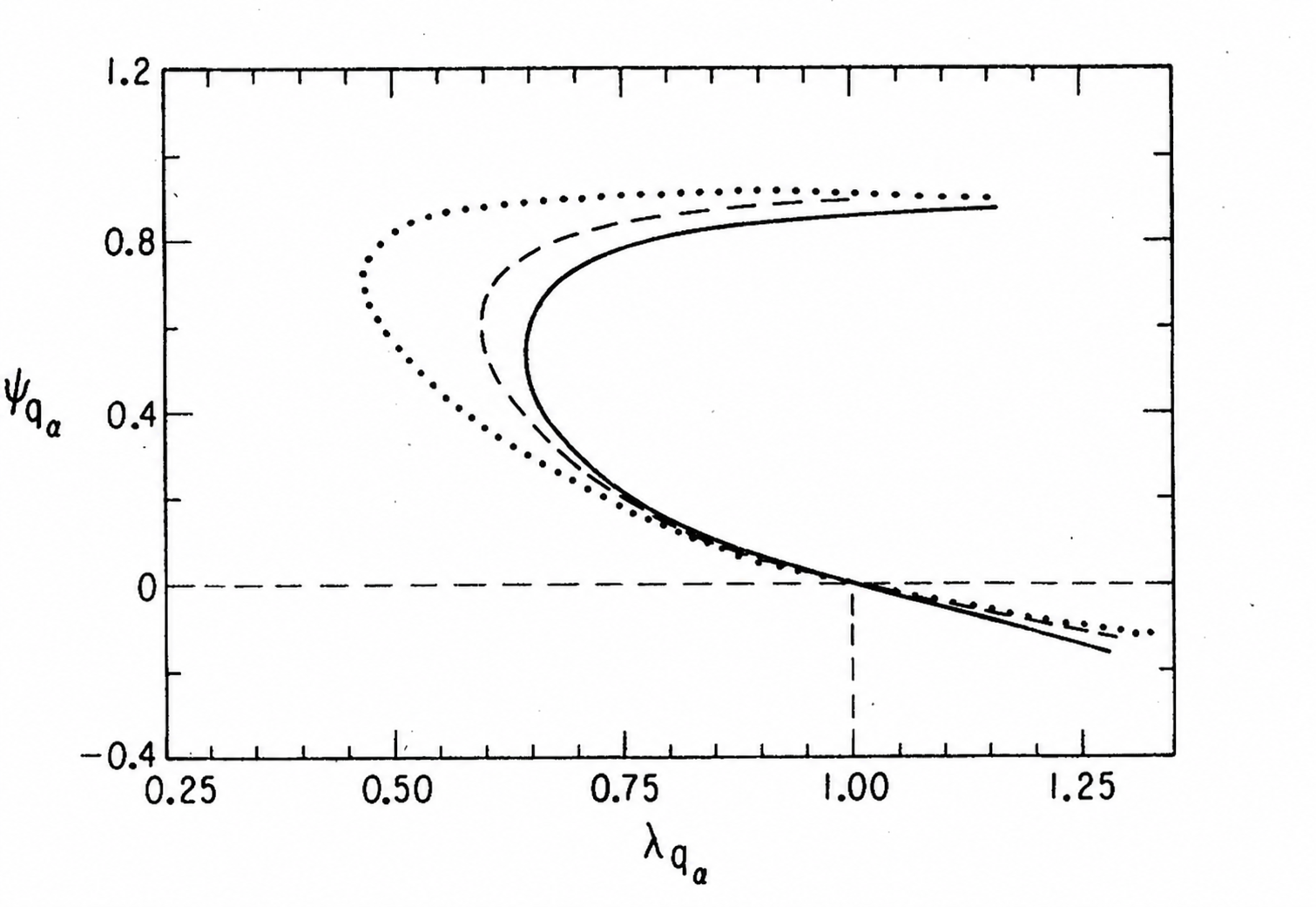}
    \caption{Three-dimensional bcc solution curves in the two-order-parameter
    theory. The dotted, dashed, and solid curves correspond to
    $\lambda_{q_\beta}=0.20$, $0.10$, and $0.05$, respectively. The common
    point $(\lambda_{q_\alpha},\psi_{q_\alpha})=(1,0)$ is marginal but is not
    an fcc-like returning bifurcation; only the lower-density freezing
    bifurcation is present. Adapted from Fig.~4 of Cerjan and Bagchi
    \cite{CerjanBagchi1985}.}
    \label{fig:bcc_CB}
\end{figure}

The bcc calculation of Cerjan and Bagchi is the closest three-dimensional counterexample to identifying $\lambda_G=1$ indiscriminately with a second bifurcation. Figure~\ref{fig:bcc_CB} shows their two-order-parameter result for the reciprocal-lattice representation used in liquid-sodium freezing \cite{CerjanBagchi1985}. The zero-amplitude solution is marginal at $(\lambda_{q_\alpha},\psi_{q_\alpha})=(1,0)$, but the diagram has only the lower-density freezing bifurcation. The fcc-like returning branch is absent.

\subsection{Two-dimensional marginal solution and the Radloff diagram}

\begin{figure}[!htbp]
    \centering
    \includegraphics[width=0.78\textwidth]{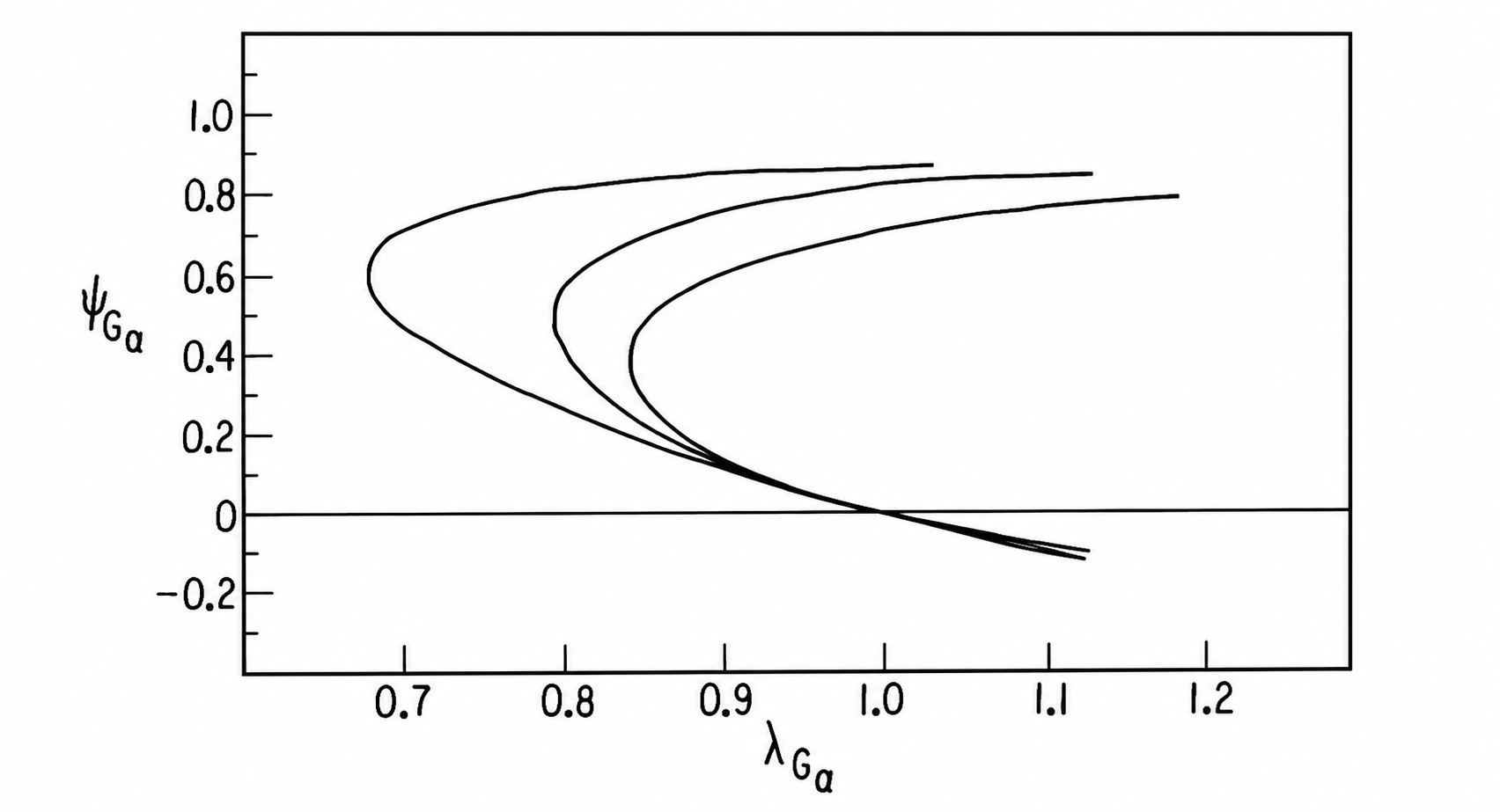}
    \caption{Two-order-parameter solution curves for a two-dimensional
    hexagonal lattice at three values of the second-mode coupling. From the
    leftmost to the rightmost turning point, the curves correspond to
    $\lambda_{G_\beta}=0.4$, $0.2$, and $0.0$, respectively. Their common
    solution $(\lambda_{G_\alpha},\psi_{G_\alpha})=(1,0)$ is marginal but is
    not the fcc-like returning bifurcation shown in
    Fig.~\ref{fig:bifurcation_3D}. Adapted from Fig.~1 of Radloff
    \textit{et al.}~\cite{Radloff1984}.}
    \label{fig:radloff_2D}
\end{figure}

The two-dimensional calculation gives the same warning. Figure~\ref{fig:radloff_2D} shows the two-order-parameter curves of Radloff \textit{et al.} for a hexagonal representation of the two-dimensional one-component plasma (OCP) \cite{Radloff1984}. Changing the second-star coupling $\lambda_{G_\beta}$ alters the global shape and turning point of the nonzero branch, while $(\lambda_{G_\alpha},\psi_{G_\alpha})=(1,0)$ remains the marginal solution.

The OCP diagram establishes the robustness of the marginal condition, not the universality of the fcc branch topology. The latter depends on lattice symmetry, mode coupling, and the lowest nonlinear invariants permitted by the reciprocal-lattice representation.

The hard-disk density is a separate two-dimensional result. Using the Baus--Colot representation of the direct correlation function, Xu and Rice located the metastable-fluid stability limit at \cite{BausColot1987,XuRice2010}
\begin{equation}
\eta_\ell^{T}\simeq0.8407,
\qquad
\phi_0=\frac{\rho_s-\rho_\ell}{\rho_\ell}\simeq0.003,
\qquad
S(G)\simeq300.
\label{eq:hard_disk_marginal_density}
\end{equation}
The density discontinuity is consequently very small in this calculation, and
\begin{equation}
\rho_s\widetilde c(G)
=(1+\phi_0)\left[1-\frac{1}{S(G)}\right]
\simeq1.
\label{eq:hard_disk_lambda_one}
\end{equation}
The numerical proximity of $\eta_\ell$ to the two-dimensional maximally random jammed regime is suggestive. It must, however, be kept logically separate from the OCP branch topology in Fig.~\ref{fig:radloff_2D}: the former concerns hard-disk terminal density, whereas the latter illustrates how marginality can survive without an fcc-like returning bifurcation.

\subsection{Coexistence, metastability, and the upper crystalline branch}

Because freezing is first order, the existence of a nonzero solution does not locate equilibrium freezing. Since $\Delta\Omega$ in Sec.~2 is defined relative to the reference liquid at the same chemical potential, coexistence requires stationary liquid and solid profiles together with
\begin{equation}
\Delta\Omega_{\mathrm{solid}}=0.
\label{eq:equal_omega}
\end{equation}

The upper crystalline branch represents progressively stronger density localization and, after the free-energy comparison selects the solid, its stable continuation. Its extension to large $|\psi_G|$ is consistent with an increasingly localized crystal, but a low-mode truncation should not be expected to describe crystalline close packing quantitatively. Stability cannot be inferred merely from whether a curve bends upward or downward.

The route relevant to the RCP-like BCR density is different. One follows the uniform liquid branch beyond equilibrium freezing as a metastable compressed liquid. Because $\psi_G=0$ remains an exact solution, this continuation exists mathematically after the liquid has ceased to be the globally stable phase. In the fcc and one-dimensional diagrams, it reaches the returning branch at
\begin{equation}
\lambda_G=1.
\label{eq:second_bifurcation}
\end{equation}

The physically relevant distinction is therefore
\begin{align}
\text{equilibrium route:}\qquad&
\text{liquid}
\longrightarrow
\text{stable crystal},
\label{eq:equilibrium_route}
\\
\text{metastable continuation:}\qquad&
\text{liquid}
\longrightarrow
\text{compressed metastable liquid}
\longrightarrow
\lambda_G=1.
\label{eq:metastable_route}
\end{align}
The second route, which is not an equilibrium freezing route, produces the RCP-like liquid density.

\subsection{The hard-sphere densities at the second bifurcation}

For hard spheres, the BCR consistency conditions give at the returning fcc bifurcation approximately \cite{BagchiCerjanRice1983}
\begin{equation}
\rho_\ell^{T}\sigma^3\simeq1.202,
\qquad
\rho_s^{T}\sigma^3\simeq1.381.
\label{eq:bcr_values_section3}
\end{equation}
The corresponding liquid packing fraction is
\begin{equation}
\eta_\ell^{T}
=
\frac{\pi}{6}\rho_\ell^{T}\sigma^3
\simeq0.629.
\label{eq:bcr_phi_section3}
\end{equation}
Thus the terminal reference-liquid packing fraction is $\eta_\ell^{T}\simeq0.629$, close to the traditional random-close-packing value, while $\rho_s^{T}\sigma^3$ lies near the crystalline close-packing density $\sqrt{2}$.

The appearance of two densities at the same bifurcation is natural in a theory with a density discontinuity:
\begin{equation}
1+\phi_0
=
\frac{\rho_s}{\rho_\ell}.
\label{eq:density_discontinuity_bcr}
\end{equation}
The liquid density characterizes the metastable reference liquid whose direct correlation function enters the theory. The solid density characterizes the average density of the inhomogeneous branch on which the crystalline density wave is defined.

The central question is therefore not simply why $\rho_\ell$ and $\rho_s$ differ, but why the marginality of the modulated branch is selected by the mixed-density combination
\begin{equation}
\rho_s\ct(G;\rho_\ell)=1.
\label{eq:central_question_end}
\end{equation}
Section~4 addresses this question directly and shows, without repeating the bifurcation construction, that the same condition is precisely the loss of RY free-energy curvature in the finite-wavevector direction.

\section{Ramakrishnan--Yussouff Curvature and the Origin of the BCR Condition}
\label{sec:curvature}

\subsection{The thermodynamic question}

The nonlinear analysis of Sec.~3 identifies the marginal condition
\begin{equation}
\lambda_G^{T}=\rho_s^{T}c_G^{T}=1.
\label{eq:curv_bcr_condition}
\end{equation}
In the fcc hard-sphere and one-dimensional hard-rod diagrams, this condition is realized where a nonzero branch returns to the uniform solution. In the two-dimensional hexagonal and three-dimensional bcc representations, by contrast, $(\lambda_G,\psi_G)=(1,0)$ remains a marginal solution but is not a second returning bifurcation. The local condition must therefore be distinguished from its symmetry-dependent nonlinear realization.

There is also a separate thermodynamic question. The existence of a stationary nonzero solution does not locate a first-order freezing transition. Because the RY grand-potential difference is defined relative to the liquid at the same temperature and chemical potential, coexistence requires
\begin{equation}
\Delta\Omega_{\mathrm{solid}}=0.
\label{eq:curv_coexistence}
\end{equation}
In the projected order-parameter description, this equality is the thermodynamic counterpart of a Maxwell tie-line construction. On the $\psi_G$--$\lambda_G$ bifurcation diagram, the tie line connects the uniform liquid branch, $\psi_G=0$, to the upper crystalline branch at the point where their grand potentials are equal. This is a particularly useful way to interpret the ordinary freezing part of the diagram. The equality cannot, however, be obtained from an equal-area rule applied directly to the plotted curve; it must be evaluated from the RY functional.

The structural coordinate on the solid side already contains the liquid--solid density change. Indeed,
\begin{equation}
\lambda_G^{(s)}
=\rho_s\widetilde c(G;\rho_\ell)
=\rho_\ell(1+\phi_0)\widetilde c(G;\rho_\ell)
=(1+\phi_0)\lambda_G^{(\ell)},
\label{eq:curv_lambda_density_jump}
\end{equation}
where $\lambda_G^{(\ell)}\equiv\rho_\ell\widetilde c(G;\rho_\ell)$ and $\phi_0=(\rho_s-\rho_\ell)/\rho_\ell$ is the fractional density discontinuity. Thus the Maxwell-like connection between the liquid and upper crystalline branches is not drawn in a density-blind order-parameter plane: the BCR coupling itself retains the fractional density change. Coexistence, marginality, and nonlinear branch topology are related but distinct statements.

The purpose of this section is to identify the thermodynamic quantity that becomes marginal at Eq.~\eqref{eq:curv_bcr_condition}. The answer is the RY curvature in a finite-wavevector density direction.

The same density-shifted factor later appeared in the nonlinear diffusion
theory of freezing \cite{Bagchi1987PhysLett,Bagchi1987Physica}.  There the
condition $1-\rho_\ell(1+\phi_0)\widetilde c(G)=0$, together with vanishing
crystalline amplitudes, makes the coupled generalized relaxation rates vanish
simultaneously.  Section~6 returns to this dynamical result; the present section
first establishes the static thermodynamic structure and its separation from
the Maxwell coexistence condition.

\subsection{One-mode Maxwell crossing and the two crystalline branches}

Before examining the terminal curvature, it is useful to make the ordinary
freezing part of the diagram quantitative.  The one-mode equation is already
sufficient for this purpose.  Introduce
\begin{equation}
t=\lambda_G\psi_G,
\qquad
I_0(t)=\avg{\exp[t\zeta_G(\br)]},
\label{eq:curv_t_I0}
\end{equation}
and normalize the periodic density at fixed mean density as
\begin{equation}
p_t(\br)=\frac{\exp[t\zeta_G(\br)]}{I_0(t)}.
\label{eq:curv_normalized_profile}
\end{equation}
For the first fcc reciprocal-lattice star,
$D_G=\avg{\zeta_G^2}=8$.  The projected stationarity equation can then be
written parametrically as
\begin{equation}
\psi_G(t)=
\frac{\avg{\zeta_G\exp(t\zeta_G)}}{D_G I_0(t)},
\qquad
\lambda_G(t)=\frac{t}{\psi_G(t)}.
\label{eq:curv_parametric_branch}
\end{equation}
The corresponding reduced grand-potential difference, evaluated relative to
the uniform state at the same mean density, is
\begin{align}
\frac{\beta\Delta\Omega_{\mathrm{1m}}}{N}
&=\avg{p_t\ln p_t}
-\frac{D_G}{2}\lambda_G\psi_G^2
\nonumber\\
&=\frac{D_G}{2}\lambda_G\psi_G^2-\ln I_0(t)
=\frac{D_G}{2}t\psi_G-\ln I_0(t).
\label{eq:curv_reduced_omega}
\end{align}
Because the two profiles compared in Eq.~\eqref{eq:curv_reduced_omega} have
the same $N$ and $V$, the chemical-potential contribution cancels.  The same
number may therefore be read as the corresponding Helmholtz free-energy
difference in this fixed-mean-density projection.  We retain the
grand-potential notation to keep the connection with the RY functional and
with the subsequent coexistence discussion explicit.
This expression reconstructs the thermodynamic landscape whose stationary
points form the one-mode bifurcation curve.  Its numerical landmarks are
listed in Table~\ref{tab:one_mode_thermodynamics}.

\begin{table}[htbp]
\centering
\normalsize
\caption{Thermodynamic landmarks of the one-mode fcc bifurcation diagram.
The grand-potential difference is measured relative to the uniform branch at
the same mean density.}
\label{tab:one_mode_thermodynamics}
\begin{tabular}{@{}lccc@{}}
\toprule
Landmark & $\lambda_G$ & $\psi_G$ & $\beta\Delta\Omega_{\mathrm{1m}}/N$ \\
\midrule
First bifurcation (fold) & 0.972991 & 0.361937 & $+0.004328$ \\
Projected Maxwell crossing & 0.979014 & 0.452229 & $0$ \\
Upper branch at $\lambda_G=1$ & 1.000000 & 0.538230 & $-0.021165$ \\
\bottomrule
\end{tabular}
\end{table}

\begin{figure}[htbp]
\centering
\includegraphics[width=0.86\textwidth]{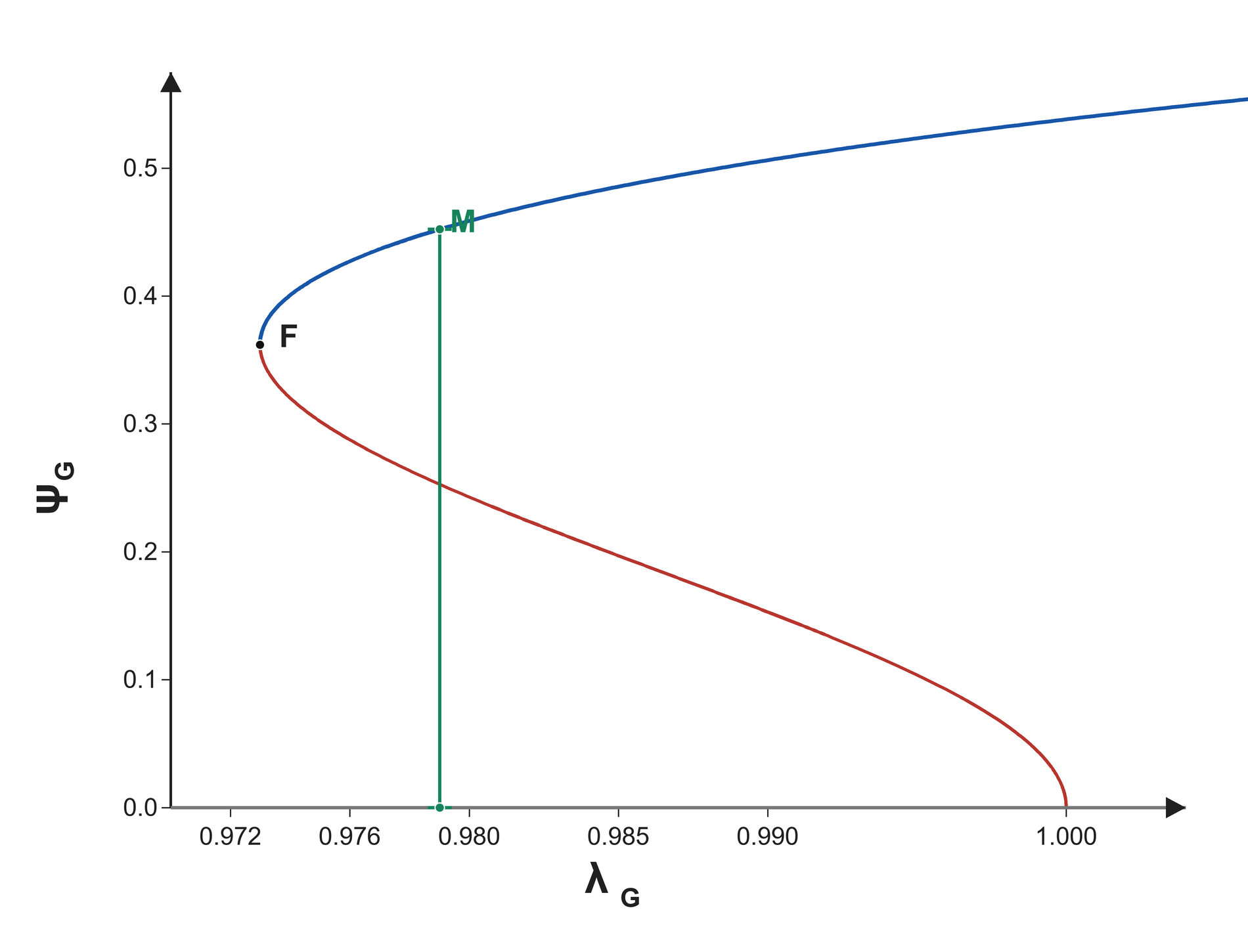}
\caption{Expanded one-mode fcc bifurcation diagram near the first
bifurcation. The blue and red curves are, respectively, the upper crystalline
minimum and the lower barrier branch; the gray horizontal line is the uniform
liquid. The branches meet at the fold $F$, where
$(\lambda_G,\psi_G)=(0.972991,0.361937)$ and the crystalline grand potential
is still higher than the liquid value by $0.004328\,Nk_{\mathrm B}T$. The
green vertical segment is the projected Maxwell tie line at
$M$, $\lambda_G=0.979014$, joining the equal-grand-potential liquid and upper
crystalline stationary states.}
\label{fig:one_mode_maxwell}
\end{figure}

The fold therefore creates the two nonuniform stationary branches while the
liquid is still the stable state.  The upper crystalline branch is born only
slightly above it, by about $0.00433\,k_{\mathrm B}T$ per particle, and is
initially metastable.  Its grand potential decreases rapidly and crosses the
liquid value at $\lambda_G\simeq0.9790$, remarkably close to the fold at
$\lambda_G\simeq0.9730$.  Beyond this crossing the upper branch is the stable
minimum of the projected landscape.

The lower branch has a different status.  From the fold to its return at
$(\lambda_G,\psi_G)=(1,0)$ its grand-potential difference remains positive
and decreases continuously from $0.004328$ to zero.  Its curvature in the
one-dimensional order-parameter direction is negative: it is the intervening
barrier, or saddle, rather than a metastable thermodynamic minimum.  The upper
branch, lower branch, and liquid branch therefore have sharply different
roles even though all three solve the same projected stationarity equation.

Equation~\eqref{eq:curv_reduced_omega} is a Maxwell construction within the
fixed-mean-density one-mode landscape.  Full liquid--solid coexistence with a
density discontinuity requires both $z_\ell=z_s$ (equal chemical potential)
and $\Omega_\ell/V_\ell=\Omega_s/V_s$; since $\Omega/V=-P$ for a homogeneous
bulk phase, the latter is equality of pressure.  The reduced crossing
should consequently not be confused with a complete two-density coexistence
calculation.  Its value is that it exposes, with no multimode complication,
the thermodynamic organization of the bifurcation diagram.

\subsection{Curvature of the one-mode functional}

We use the one-mode representation and free energy already obtained in Sec.~2. With
\begin{equation}
A(\br)=1+\phi_0+u_G\zeta_G(\br),
\qquad
\rho_s=\rho_\ell(1+\phi_0),
\label{eq:curv_A_definition}
\end{equation}
the density is $\rho(\br)=\rho_\ell A(\br)$, while $\avg{\zeta_G}=0$ and $D_G=\avg{\zeta_G^2}$. The relevant one-mode grand-potential density is
\begin{align}
\frac{\beta\Delta\Omega}{V}
={}&\rho_\ell\avg{A\ln A-(A-1)}
\nonumber\\
&-\frac{\rho_\ell^2}{2}
\left[\widetilde c(0;\rho_\ell)\phi_0^2
+D_G\widetilde c(G;\rho_\ell)u_G^2\right].
\label{eq:curv_one_mode_free_energy}
\end{align}
This single expression contains everything needed for the marginality calculation.

At fixed $\phi_0$, differentiation of the ideal term gives
\begin{equation}
\frac{\partial^2}{\partial u_G^2}
\left(\frac{\beta\Delta\Omega_{\mathrm{id}}}{V}\right)
=\rho_\ell\avg{\frac{\zeta_G^2}{A}}.
\label{eq:curv_ideal_general}
\end{equation}
At $u_G=0$, the profile is the density-shifted uniform endpoint associated with the modulated-branch parametrization:
\begin{equation}
A=1+\phi_0=\frac{\rho_s}{\rho_\ell}.
\label{eq:curv_shifted_endpoint}
\end{equation}
Consequently, the ideal entropy contributes the positive curvature
\begin{equation}
K_{\mathrm{id}}
=\left.
\frac{\partial^2(\beta\Delta\Omega_{\mathrm{id}}/V)}
{\partial u_G^2}
\right|_{u_G=0}
=\frac{\rho_\ell^2D_G}{\rho_s}.
\label{eq:curv_ideal_result}
\end{equation}
The factor $1/\rho_s$ is therefore not introduced phenomenologically. It comes from evaluating the ideal entropy curvature about the density-shifted endpoint rather than about the original reference liquid.

The finite-$G$ excess term is already quadratic, and hence
\begin{equation}
K_{\mathrm{ex}}
=\left.
\frac{\partial^2(\beta\Delta\Omega_{\mathrm{ex}}/V)}
{\partial u_G^2}
\right|_{u_G=0}
=-\rho_\ell^2D_G\widetilde c(G;\rho_\ell).
\label{eq:curv_excess_result}
\end{equation}
The ideal part resists density modulation, whereas the correlation contribution favors it when $\widetilde c(G;\rho_\ell)>0$.

\subsection{The BCR condition as finite-wavevector marginality}

Adding the two curvatures gives the central result
\begin{equation}
K_G\equiv
\left.
\frac{\partial^2(\beta\Delta\Omega/V)}
{\partial u_G^2}
\right|_{u_G=0,\,\phi_0}
=\rho_\ell^2D_G
\left[\frac{1}{\rho_s}-\widetilde c(G;\rho_\ell)\right].
\label{eq:curv_main_result}
\end{equation}
Thus
\begin{equation}
K_G=0
\quad\Longleftrightarrow\quad
\rho_s c_G=1
\quad\Longleftrightarrow\quad
\lambda_G=1.
\label{eq:curv_equivalence}
\end{equation}
The condition obtained from the nonlinear BCR equation is therefore exactly the condition for loss of quadratic RY stiffness in the finite-$G$ direction. For $\lambda_G<1$ the density-shifted state resists the modulation, at $\lambda_G=1$ it is marginal, and for $\lambda_G>1$ the projected curvature is negative.

This is not the ordinary liquid instability $\rho_\ell\widetilde c(G;\rho_\ell)=1$, which would make the liquid structure factor diverge. The BCR condition instead mixes the direct correlation function of the metastable reference liquid at $\rho_\ell$ with the mean density $\rho_s$ associated with the modulated branch. The mixed-density structure is precisely what allows marginality to occur while the liquid peak $S(G)$ is large but finite.

\subsection{Why the projected curvature is meaningful}

Equation~\eqref{eq:curv_main_result} is evaluated at fixed $\phi_0$, but the finite-$G$ direction is not arbitrarily isolated at the endpoint. The mixed Hessian element vanishes there:
\begin{equation}
H_{\phi_0 u_G}
=\left.
\frac{\partial^2(\beta\Delta\Omega/V)}
{\partial\phi_0\,\partial u_G}
\right|_{u_G=0}
=0,
\label{eq:curv_mixed_hessian}
\end{equation}
because every mixed contribution is proportional to $\avg{\zeta_G}=0$. The uniform density shift and the finite-$G$ modulation therefore decouple to quadratic order at the uniform endpoint. Away from that endpoint, or when other nonzero reciprocal-lattice modes are retained, the complete Hessian must be considered.

This quadratic decoupling does not assert that the density-shifted profile at
$u_G=0$ is automatically a second equilibrium liquid. It is the limiting
profile inherited from the inhomogeneous branch. Stationarity with respect to
the uniform density coordinate, and global comparison with the reference
liquid, remain separate requirements. The curvature result concerns the
finite-$G$ direction at that specified branch endpoint.

The common factor $D_G$ also has a simple meaning. Rescaling the lattice function changes $D_G$ and rescales both $u_G$ and $\psi_G$ inversely, but it does not move the zero of $K_G$. The marginal condition is consequently independent of the arbitrary normalization chosen for the reciprocal-lattice basis. This is a coordinate-invariance property of the one-mode RY construction, not a claim of full thermodynamic universality.

\subsection{What the result establishes}

The result establishes an exact identity within the one-mode RY/BCR framework: the BCR condition is a local finite-wavevector thermodynamic marginality. It also explains why $\rho_s$, rather than $\rho_\ell$, enters the bifurcation parameter. The ideal entropy supplies $1/\rho_s$, the excess functional supplies $-\widetilde c(G;\rho_\ell)$, and their cancellation leaves Eq.~\eqref{eq:curv_bcr_condition}.

Three distinctions remain essential. First, zero curvature does not determine equilibrium freezing; the latter is selected by the coexistence condition in Eq.~\eqref{eq:curv_coexistence}, equivalent to the Maxwell-like equality of the two stationary grand-potential branches. Second, thermodynamic marginality does not by itself establish mechanical jamming: the RY theory contains no contact-force network, rigidity matrix, or isostaticity condition. Third, the local marginal condition does not determine the global nonlinear topology. The fcc and one-dimensional representations realize it as a returning bifurcation, whereas the two-dimensional hexagonal and three-dimensional bcc representations do not.

The robustness of the algebraic cancellation must therefore be separated from the numerical and topological questions. Its extension to several coupled reciprocal-lattice stars and to more complete free-energy functionals requires examination of the full Hessian. Most importantly, the curvature calculation alone does not explain why the resulting density is exact terminal packing in one dimension and lies close to random terminal packing in two and three dimensions. The evidence and its limitations are collected in Table~\ref{tab:dimensional_comparison}.

\begin{table}[htbp]
\centering
\normalsize
\caption{Dimensional comparison of the terminal condition $\lambda_G^{T}=\rho_s^{T}c_G^{T}=1$. The two-dimensional density and topology entries come from different systems and should not be conflated: the density estimate is for hard disks, whereas the displayed hexagonal branch topology is for the two-dimensional OCP.}
\label{tab:dimensional_comparison}
\begin{tabular}{@{}>{\raggedright\arraybackslash}p{0.17\textwidth}>{\raggedright\arraybackslash}p{0.27\textwidth}>{\raggedright\arraybackslash}p{0.47\textwidth}@{}}
\toprule
Dimension and system & Quantitative result & Topology and interpretation \\
\midrule
1D hard rods & $\rho_\ell^{T}\sigma=\rho_s^{T}\sigma=1$ & The only bifurcation: the modulated branch meets the uniform branch at $\lambda_G^{T}=1$. This is exact terminal packing, although hard rods have no equilibrium freezing transition. \\
\addlinespace
2D hard disks / hexagonal OCP & Hard disks: $\eta_\ell^{T}\simeq0.8407$, $\phi_0^{T}\simeq0.003$ & The hard-disk density lies near the maximally random jammed regime. In the separate hexagonal-OCP calculation, $(\lambda_G,\psi_G)=(1,0)$ is marginal but is not an fcc-like returning bifurcation. \\
\addlinespace
3D hard spheres & fcc: $\eta_\ell^{T}\simeq0.629$, $\rho_s^{T}\sigma^3\simeq1.381$ & The fcc representation has a returning branch and selects a liquid density near traditional RCP. The bcc representation retains the marginal solution without a second returning bifurcation; topology depends on lattice symmetry. \\
\bottomrule
\end{tabular}
\end{table}

That deeper structural puzzle is taken up after the local stability analysis.

\section{Local Equivalence of the Cerjan--Bagchi Criterion and RY Curvature}
\label{sec:CB_RY_stability}

\subsection{Residual formulation of the BCR equation}

Section~3 expressed the one-mode BCR equation as
\begin{equation}
D_G\psi_G=F_G(\lambda_G\psi_G),
\qquad
F_G(x)=
\frac{\avg{\zeta_Ge^{x\zeta_G}}}
{\avg{e^{x\zeta_G}}}.
\label{eq:cb_bcr_equation}
\end{equation}
It is useful to collect the difference between the assumed and generated density-wave amplitudes in the residual
\begin{equation}
R_G(\psi_G,\lambda_G)
=D_G\psi_G-F_G(\lambda_G\psi_G).
\label{eq:cb_residual}
\end{equation}
A solution branch satisfies $R_G=0$. The derivative $\partial R_G/\partial\psi_G$ measures the linear response of this self-consistency condition to a displacement from the branch; it is a branch stiffness, not yet by itself a thermodynamic curvature.

The nonlinear response has a useful statistical interpretation. If averages are reweighted by $e^{x\zeta_G}$, then $F_G(x)=\langle\zeta_G\rangle_x$ and
\begin{equation}
F_G'(x)
=\langle\zeta_G^2\rangle_x
-\langle\zeta_G\rangle_x^2
=\operatorname{Var}_x(\zeta_G)>0.
\label{eq:cb_variance}
\end{equation}
At the uniform endpoint, $x=0$, so $F_G'(0)=\avg{\zeta_G^2}=D_G$. Therefore,
\begin{equation}
\left.
\frac{\partial R_G}{\partial\psi_G}
\right|_{\psi_G=0}
=D_G(1-\lambda_G).
\label{eq:cb_residual_endpoint}
\end{equation}
The BCR equation consequently loses linear stiffness at $\lambda_G=1$. This statement is local and does not assume that the nonlinear branch forms an fcc-like returning bifurcation.

\subsection{Bridge to the RY curvature}

The RY calculation of Sec.~4 gave, at the same density-shifted uniform endpoint,
\begin{equation}
K_G^{\mathrm{RY}}
=\left.
\frac{\partial^2(\beta\Delta\Omega/V)}
{\partial u_G^2}
\right|_{u_G=0,\,\phi_0}
=\frac{\rho_\ell^2D_G}{\rho_s}(1-\lambda_G).
\label{eq:cb_RY_curvature}
\end{equation}
Comparison with Eq.~\eqref{eq:cb_residual_endpoint} yields the central result
\begin{equation}
K_G^{\mathrm{RY}}
=\frac{\rho_\ell^2}{\rho_s}
\left.
\frac{\partial R_G}{\partial\psi_G}
\right|_{\psi_G=0}.
\label{eq:cb_bridge}
\end{equation}
The prefactor is strictly positive and follows from $u_G=(\rho_s/\rho_\ell)\psi_G$. Hence the Cerjan--Bagchi residual stiffness and the RY finite-$G$ curvature have the same sign and vanish at the same point. The earlier bifurcation-theory stability factor is therefore the local representation of an RY thermodynamic curvature at the uniform endpoint.

The qualification ``at the uniform endpoint'' is essential. Equation~\eqref{eq:cb_bridge} does not prove that the residual derivative equals the full thermodynamic Hessian everywhere on a nonlinear branch. Away from $\psi_G=0$, other modes and the uniform density shift may couple, and the relation between the self-consistency residual and the free-energy curvature must be established for the full parametrization.

\subsection{Branch slope and turning points}

The geometry of a nonzero solution branch still contains useful local information. Implicit differentiation of $R_G(\psi_G,\lambda_G)=0$ gives
\begin{equation}
\frac{d\psi_G}{d\lambda_G}
=\frac{\psi_GF_G'(\lambda_G\psi_G)}
{D_G-\lambda_GF_G'(\lambda_G\psi_G)}.
\label{eq:cb_branch_slope}
\end{equation}
For the convention $\psi_G>0$, the numerator is positive because $F_G'>0$. A vertical tangent therefore occurs when
\begin{equation}
D_G-\lambda_GF_G'(\lambda_G\psi_G)
=\frac{\partial R_G}{\partial\psi_G}=0.
\label{eq:cb_turning_point}
\end{equation}
This explains why turning points were useful indicators in the Cerjan--Bagchi analysis: they mark a singularity of the linearized branch equation.

The slope must nevertheless be interpreted as a coordinate-dependent geometric diagnostic. Its sign changes if the order-parameter convention is reversed, and $\lambda_G$ is not an externally imposed pressure, volume, or chemical potential. A bend in the $\psi_G$--$\lambda_G$ curve is therefore neither a compressibility nor, by itself, proof of thermodynamic instability. At the uniform endpoint, Eq.~\eqref{eq:cb_bridge} supplies the required thermodynamic interpretation; elsewhere, stability requires the appropriate free-energy Hessian.

\subsection{Local stability, coexistence, and the physical control variables}

The local criterion and the global phase selection answer different questions. Positive $K_G^{\mathrm{RY}}$ means that an infinitesimal finite-$G$ modulation raises the grand potential; zero curvature means marginality; and negative curvature means instability in that projected direction. Global stability instead requires comparison of grand-potential values. At freezing, the Maxwell-like tie line discussed in Sec.~4 connects the uniform liquid and upper crystalline branches when
\begin{equation}
\Delta\Omega_{\mathrm{solid}}=0.
\label{eq:cb_coexistence}
\end{equation}
Thus neither the residual sign nor the slope of the bifurcation curve locates coexistence without evaluation of the thermodynamic potential.

The horizontal coordinate of the bifurcation diagram is itself a derived structural quantity:
\begin{equation}
\lambda_G
=\rho_\ell(1+\phi_0)
\widetilde c\!\left(G(\rho_s);\rho_\ell\right).
\label{eq:cb_composite_lambda}
\end{equation}
It combines the direct correlation function of the reference liquid, the fractional density discontinuity, and the reciprocal-lattice scale of the solid. The diagram therefore organizes nonlinear solutions and their local singularities; the thermodynamic embedding is supplied by the RY functional and the physical control variables $\rho_\ell$, pressure, and chemical potential.

Within this precise local domain, the Cerjan--Bagchi and RY analyses are not competing descriptions. They identify the same finite-wavevector marginality through, respectively, the Jacobian of the nonlinear self-consistency equation and the Hessian of the thermodynamic potential. This equivalence explains the stability meaning of $\lambda_G^{T}=1$, but not why that terminal condition selects packing exactly in one dimension and approximately in two and three dimensions. That is the structural question addressed next.

\section{Interpretation and Scope of the BCR Endpoint}
\label{sec:rcp_interpretation}

\subsection{Finite-wavevector marginality and its thermodynamic scope}

The preceding sections establish the local thermodynamic meaning of the BCR
endpoint.  The condition
\begin{equation}
\lambda_G^{T}=\rho_s^{T}c_G^{T}=1
\label{eq:sec6_marginal_condition}
\end{equation}
is the zero of the RY curvature for a finite-wavevector density modulation.
The Cerjan--Bagchi residual loses its linear stiffness at the same point, so
the result is not an accidental root of the projected nonlinear equation.  In
the fcc representation the returning nonzero branch also reaches the uniform
solution there.  The topology is less general than the marginality: bcc,
simple-cubic, and two-dimensional hexagonal representations may contain the
solution $(\lambda_G,\psi_G)=(1,0)$ without an fcc-like returning branch.

The vanishing curvature invites comparison with a spinodal, but the soft
direction has $G\neq0$ and $\delta\rho(\mathbf r)\propto\zeta_G(\mathbf r)$,
not the long-wavelength density mode of a liquid--gas spinodal.  The endpoint
is therefore most accurately described as a finite-$G$ thermodynamic
marginality.  The returning branch is the lower barrier branch identified in
Sec.~4, whereas the stable crystalline branch continues toward stronger
localization.  Neither a divergent compressibility nor a dynamical or
mechanical instability follows from the static RY calculation.

This distinction also separates the endpoint from equilibrium freezing.  The
one-mode Maxwell crossing occurs on the upper branch before $\lambda_G=1$,
whereas the endpoint concerns the loss of local stiffness of a different
stationary branch.  Coexistence, branch stability, and terminal marginality
are therefore three related but distinct properties of the same nonlinear
landscape.

\subsection{Retained density, structural response, and dimensional constraint}

The endpoint has the mixed character
\begin{equation}
\psi_G^{T}=0,
\qquad
\phi_0^{T}=\frac{\rho_s^{T}-\rho_\ell^{T}}{\rho_\ell^{T}},
\qquad
\rho_s^{T}=\rho_\ell^{T}(1+\phi_0^{T}).
\label{eq:sec6_mixed_endpoint}
\end{equation}
The periodic amplitude disappears, but the mean-density scale carried by the
inhomogeneous branch remains distinct in the two- and three-dimensional
calculations, for which $\phi_0$ is nonzero.  The exact one-dimensional endpoint
is the important exception: there $\rho_\ell=\rho_s$ and $\phi_0=0$.  In the
higher-dimensional cases the endpoint is consequently not a conventional
continuous transition in which the liquid and inhomogeneous branches become
identical in all their thermodynamic coordinates.  At the level of the
one-body density, $\psi_G=0$ establishes only the loss of crystalline contrast.
It does not identify an amorphous solid; such a distinction would require
higher-order correlations, localization, or broken ergodicity.

The density is selected through
\begin{equation}
\rho_s^{T}c_G^{T}=1,
\qquad
c_G^{T}=\widetilde c(G;\rho_\ell^{T})=\frac{1}{\rho_s^{T}}.
\label{eq:sec6_structural_balance}
\end{equation}
Here $\rho_s$ supplies the density scale of the inhomogeneous branch, while
$\widetilde c(G;\rho_\ell)$ measures the response of the dense reference
liquid at the selected packing wavelength.  The balance combines the ideal
entropic stiffness, the finite-$G$ correlation response, and the retained
density shift.  It may therefore register saturation of the liquid's ability
to sustain an additional density wave.  This remains an interpretation rather
than a derivation because the response is that of a uniform reference liquid
and the mapping from $G$ to density retains lattice information.

The strongest constraint on this interpretation is dimensional.  Bagchi and
Rice showed that the two-body density-functional representation used here is
exact both for $d=1$ and in the limit $d\rightarrow\infty$
\cite{BagchiRice1988}.  In one dimension, for hard rods,
\begin{equation}
\rho_\ell^{T}\sigma=\rho_s^{T}\sigma=1
\label{eq:sec6_1d_close_packing}
\end{equation}
is both the unique bifurcation point and exact terminal packing.  The result
shows that the marginal condition can encode complete excluded-volume
saturation and is therefore evidence against a purely numerical coincidence.

The opposite limit is equally instructive.  For simple-hypercubic ordering as
$d\rightarrow\infty$, Bagchi and Rice found that the bifurcation topology is
dimensionally persistent and argued that the continuous point at
$\lambda_G=1$ selects the maximum achievable packing of that lattice family.
At the same time, the effective harmonic restoring force of the emerging
hypercubic solid tends to zero in this limit.  The continuous bifurcation thus
marks a limiting packing condition without implying the appearance of a
stable ordered phase.  This distinction is important: terminal geometrical
selection and thermodynamic stabilization of a crystal are not the same
statement.

The two exact limits therefore constrain, but do not uniquely determine, the
interpretation at intermediate dimension.  In $d=1$, marginality coincides
with complete filling of the line.  At large $d$, it again selects a maximum
packing condition, although the associated ordered state is mechanically
soft.  In two and three dimensions geometrical frustration may cause the
corresponding exhaustion of accessible configurations while the packing is
still disordered and below crystalline close packing.  This offers a
nonaccidental rationale for the proximity to maximally random packing, but it
is not yet a derivation of RCP from the direct correlation function.

\subsection{Metastable continuation, limitations, and discriminating tests}

The equilibrium hard-sphere liquid freezes before reaching the BCR endpoint.
If crystallization is avoided, the relevant sequence is
\begin{equation}
\text{stable liquid}
\longrightarrow
\text{metastable liquid}
\longrightarrow
\text{finite-$G$ marginality}.
\label{eq:sec6_metastable_sequence}
\end{equation}
The endpoint is thus a limit of a metastable continuation, not an equilibrium
phase boundary.  There is, however, a direct dynamical counterpart within the
same mean-field description.  In the nonlinear diffusion theory of freezing,
the coupled reciprocal-lattice amplitudes possess generalized relaxation
rates whose common vanishing condition is
\begin{equation}
1-\rho_\ell(1+\phi_0)\widetilde c(G;\rho_\ell)=0,
\qquad \psi_G=0,
\label{eq:sec6_dynamic_condition}
\end{equation}
or, equivalently, $\rho_s^{T}c_G^{T}=1$
\cite{Bagchi1987PhysLett,Bagchi1987Physica}.  Thus the static zero of the RY
curvature and the simultaneous dynamical slowing of the retained density-wave
modes occur at the same endpoint.  The high-density hard-sphere solution found
in that work, near $\rho_\ell\sigma^3\simeq1.20$ with
$\phi_0\simeq0.15$, was originally set aside as inaccessible from the
equilibrium liquid.  In the present bifurcation diagram that inaccessibility
has a natural meaning: the solution belongs to a deeply metastable
continuation beyond ordinary freezing.

This dynamical correspondence strengthens the interpretation of
$\lambda_G^{T}=1$ as a terminal loss of restoring force, but it must not be
overstated.  The nonlinear diffusion equation describes relaxation of smooth
density modes; it contains neither a contact-force network nor a Maxwell
rigidity criterion.  It therefore does not prove that the endpoint is the
jammed RCP state.  Moreover, activated crystallization, glassy arrest, or
jamming may pre-empt the mean-field endpoint, as emphasized by modern
mean-field and finite-dimensional analyses of glass and jamming transitions
\cite{CharbonneauEtAl2017}; the physical fluid therefore need not reach the
predicted divergence of the collective relaxation time along an ergodic path.

Several tests can distinguish a structural principle from a closure- or
lattice-dependent coincidence:
\begin{enumerate}
\item repeat the density mapping using accurate alternative hard-sphere direct correlation functions;
\item retain coupling among reciprocal-lattice stars in a fully multimode RY calculation;
\item compare lattice and amorphous density representations using the same thermodynamic input; and
\item test the dimensional trend and compare the predicted common slowing of the reciprocal-lattice modes with relaxation, contact, and rigidity observables.
\end{enumerate}

The established result is precise: $\lambda_G=1$ is the zero of the finite-$G$
RY curvature, and the Cerjan--Bagchi branch-stability factor has the same sign.
The retained density shift, the structural balance in
Eq.~\eqref{eq:sec6_structural_balance}, the two exact dimensional limits, and
the independent dynamical condition in Eq.~\eqref{eq:sec6_dynamic_condition}
are mutually consistent with a terminal packing constraint encoded in the
direct correlation function.  They do not yet explain why the
three-dimensional density lies near RCP or establish mechanical jamming.
That unresolved mapping, rather than a claimed identity with RCP, is the
principal conclusion of this section.

\section{Contact Structure and Finite-Wavevector Response Near the BCR Endpoint}
\label{sec:contact_response}

\subsection{Compressibility and finite-wavelength structural response}

For an equilibrated hard-sphere fluid, the long-wavelength density response is
fixed by the compressibility sum rule,
\begin{equation}
S(0)=\rho_\ell k_{\mathrm B}T\kappa_T
=\frac{1}{1-\rho_\ell\widetilde c(0)}.
\label{eq:sec7_compressibility}
\end{equation}
Metastable compression strongly suppresses volume fluctuations, so $S(0)$
decreases and $\rho_\ell\widetilde c(0)$ becomes increasingly negative.  In
an ideally jammed state the corresponding fluctuations are no longer ordinary
equilibrium fluctuations: a volume-changing displacement can open contacts,
and rigidity depends on whether the contact constraints remain satisfied.
Thus the equilibrium sum rule motivates the approach to incompressibility but
does not, by itself, constitute a mechanical jamming criterion.

The response near the first maximum of the structure factor is different.  At
$k_m\neq0$,
\begin{equation}
S(k_m)=\frac{1}{1-\rho_\ell\widetilde c(k_m)},
\label{eq:sec7_finite_k_response}
\end{equation}
and $\widetilde c(k_m)$ describes the free-energy response to a modulation on
the scale of the first coordination shell.  There is no exact compressibility
sum rule connecting $\widetilde c(k_m)$ to $S(0)$.  Nevertheless, both are
Fourier components of the same direct correlation function, so the progressive
quenching of volume fluctuations and the reorganization of the near-neighbor
shell need not be structurally independent.

This distinction is particularly relevant to the BCR condition,
\begin{equation}
1=\rho_s\widetilde c(G;\rho_\ell)
=\frac{\rho_s}{\rho_\ell}
\left[\rho_\ell\widetilde c(G;\rho_\ell)\right].
\label{eq:sec7_mixed_density_response}
\end{equation}
Because $\rho_s/\rho_\ell>1$, Eq.~\eqref{eq:sec7_mixed_density_response} can
hold while $\rho_\ell\widetilde c(G)<1$ and $S(G)$ remains finite.  The BCR
endpoint is therefore not a divergence of the liquid structure factor.  It is
a cancellation of the finite-$G$ stiffness in a mixed-density description,
occurring against a background of increasing long-wavelength incompressibility.

\subsection{Near-contact population as a structural diagnostic}

The first peak of $S(k)$ is a collective measure and cannot by itself count
near contacts.  The hard-sphere virial relation
\begin{equation}
\frac{\beta P}{\rho_\ell}=1+4\eta_\ell g(\sigma^+)
\label{eq:sec7_virial_contact}
\end{equation}
shows that the rapid pressure growth under metastable compression is
accompanied by increasing contact crowding.  A useful measure that does not
identify jamming merely with the value at one point is the integrated
near-contact population
\begin{equation}
{\cal N}(\Delta)=4\pi\rho_\ell
\int_\sigma^{\sigma+\Delta}r^2g(r)\,dr .
\label{eq:sec7_near_contact_population}
\end{equation}
For a regular dense liquid with finite $g(\sigma^+)$,
${\cal N}(\Delta)\sim\Delta$.  Nearly jammed disordered packings instead show
an anomalous small-gap contribution
\begin{equation}
g_{\mathrm{nc}}(r)\sim(r-\sigma)^{-\gamma},
\qquad
{\cal N}_{\mathrm{nc}}(\Delta)\sim\Delta^{1-\gamma},
\qquad 0<\gamma<1,
\label{eq:sec7_near_contact_scaling}
\end{equation}
with $\gamma$ close to $0.4$ in three-dimensional disordered packings
\cite{DonevTorquatoStillinger2005}.  The exponent is not predicted by BCR or
the RY functional.  Its role here is diagnostic: a crossover from regular to
anomalous near-contact scaling near the BCR density would supply information
that the value of $S(k_m)$ alone cannot provide.

The dimensional form of this diagnostic is especially revealing.  For hard
hyperspheres of diameter $\sigma$ in $d$ dimensions, define
\begin{equation}
 {\cal N}_d(\Delta)=\rho_\ell S_{d-1}
 \int_\sigma^{\sigma+\Delta}r^{d-1}g(r)\,dr,
\label{eq:sec7_near_contact_d}
\end{equation}
where $S_{d-1}$ is the area of the unit $(d-1)$-sphere.  Introduce the scaled
gap $y=d(r-\sigma)/\sigma$ and take a shell
$\Delta=\sigma\widehat\Delta/d$.  Since
$(1+y/d)^{d-1}\rightarrow e^y$ and
$\rho_\ell S_{d-1}\sigma^d/d=2^d\eta_\ell$, Eq.~\eqref{eq:sec7_near_contact_d}
becomes
\begin{equation}
 {\cal N}_d\!\left(\frac{\sigma\widehat\Delta}{d}\right)
 \simeq 2^d\eta_\ell
 \int_0^{\widehat\Delta}e^y
 g\!\left[\sigma\left(1+\frac{y}{d}\right)\right]dy .
\label{eq:sec7_large_d_contact_shell}
\end{equation}
Thus the natural near-contact layer becomes radially narrow as $1/d$, while
its phase-space weight varies by an order-one exponential factor across the
scaled gap.  This is precisely the regime in which Bagchi--Rice dimensional
continuation is most informative \cite{BagchiRice1988}: a terminal-packing
condition may be visible in a very thin real-space shell even though its
Fourier signature is distributed across the first-peak region.

Equation~\eqref{eq:sec7_large_d_contact_shell} also suggests a sharp test.  A
regular fluid shell and an accumulating quasi-contact shell have different
limits as $d$ grows.  If the cumulative population in a window of width
$\sigma/d$ becomes of order $d$, it reaches the same dimensional scale as the
isostatic contact number $Z_c=2d$; equality is neither assumed nor implied,
because ${\cal N}_d$ counts near neighbors within a finite gap whereas $Z_c$
counts force-bearing contacts.  Studying the order of limits
$d\rightarrow\infty$ and $\widehat\Delta\rightarrow0$ could therefore reveal
whether BCR marginality is accompanied by a genuine concentration of pair
probability at contact or only by smooth coordination-shell crowding.

Contact formation affects a broad range of wavevectors rather than a single
Fourier component.  Numerical studies of nearly jammed packings find that the
direct correlation function is sensitive both to long-range suppression of
density fluctuations and to protocol-dependent contact formation
\cite{AtkinsonStillingerTorquato2016}.  A useful comparison should therefore
track $S(0)$, $\widetilde c(k)$ across the first peak, and
${\cal N}_d(\sigma\widehat\Delta/d)$ simultaneously as the BCR density is
approached.

\subsection{Relation to contact rigidity and the BCR endpoint}

Maxwell rigidity is a statement about constraints.  For frictionless spheres,
an isostatic packing has just enough independent contacts to constrain the
nontrivial displacement degrees of freedom.  A volume-changing or nonaffine
fluctuation can destroy contacts unless it is opposed by this network.  Such
information is contained in the contact geometry and mechanical Hessian, not
in the scalar height of the first peak of $S(k)$.

The BCR curvature probes the thermodynamic cost of a modulation at
approximately the nearest-neighbor wavelength, whereas Maxwell rigidity probes
the rank of the contact constraints.  The possible bridge is therefore not
the scalar height of the first peak.  It is the joint evolution of three
sectors: suppression of the $k\rightarrow0$ density response, loss of the
projected finite-$G$ restoring cost, and concentration of pair probability in
a shrinking near-contact shell.  These are complementary structural signals,
not equivalent definitions of jamming.

The proposal can be tested without assuming that equivalence.  At the density
selected by the BCR condition one should examine whether (i) $S(0)$ is strongly
suppressed, (ii) $\widetilde c(k)$ develops a distinctive change across the
first-peak region, (iii) the near-contact population crosses toward jammed-like
scaling, and (iv) contact number or rigidity diagnostics approach their
isostatic values.  Agreement of these independently defined quantities across
dimensions, closures, lattice representations, and compression protocols
would turn the present structural correspondence into a substantive
explanation of the RCP-like density.  Section~6 established that the BCR
endpoint has both static and mean-field dynamical marginality.  The present
section identifies the additional real-space evidence required to connect
that endpoint to contact rigidity.  This distinction provides the appropriate
starting point for the Conclusion.

\section{Conclusion}

This work separates three features of the density-functional bifurcation
diagram that are easily conflated: the first crystalline fold, thermodynamic
coexistence, and the terminal marginal point.  The terminal condition is
\begin{equation}
 \lambda_G^{T}=\rho_s^{T}c_G^{T}=1,
\end{equation}
where the direct correlation function is evaluated in the reference liquid at
$\rho_\ell$, while $\rho_s$ is the mean density retained by the inhomogeneous
branch.  Within the one-mode RY/BCR construction, this condition is exactly
the zero of the grand-potential curvature in the principal finite-$G$ density
direction.  The Cerjan--Bagchi residual loses its linear stiffness at the same
point.  The result is local: it identifies a thermodynamic marginality, not a
global phase boundary.

The ordinary freezing part of the bifurcation diagram must be interpreted
separately.  The first fold is the point at which two finite-amplitude
crystalline stationary solutions appear; it is not itself the thermodynamic
transition.  In the one-mode fcc calculation the fold occurs at
$\lambda_G=0.972991$, where the crystalline solution is still slightly less
stable than the liquid, $\beta\Delta\Omega/N=0.004328$.  Equality of the grand
potentials occurs only at the nearby projected Maxwell crossing,
$\lambda_G=0.979014$, on the upper branch.  This fixed-mean-density crossing
clarifies the organization of the stationary branches, but it is not a full
two-density coexistence calculation; equality of chemical potential and
pressure remains the complete thermodynamic requirement.

The dimensional evidence establishes both the strength and the limits of the
terminal result.  For one-dimensional hard rods, $\lambda_G^{T}=1$ gives
$\rho_\ell^{T}\sigma=\rho_s^{T}\sigma=1$: the only bifurcation point is exactly the
close-packed state.  For two-dimensional hard disks, the same structural
condition gives $\eta_\ell^{T}\simeq0.8407$, close to the maximally random jammed
regime.  The separate hexagonal OCP calculation shows, however, that
$(\lambda_G,\psi_G)=(1,0)$ can remain marginal without becoming a second
returning bifurcation.  In three-dimensional hard spheres, the fcc branch does
return to the uniform solution at $\lambda_G=1$, selecting
$\eta_\ell^{T}\simeq0.629$, close to the traditional RCP density, whereas the bcc
calculation retains the marginal solution without the same returning
topology.  The $d$-dimensional formulation is exact at $d=1$ and as
$d\rightarrow\infty$, but the global nonlinear topology remains dependent on
lattice symmetry and mode coupling.

The returning fcc endpoint belongs to the lower, barrier-like branch, not to
the stable crystalline branch, which continues toward stronger localization.
The earlier nonlinear diffusion theory gives the same density-shifted
condition for simultaneous vanishing of the coupled generalized relaxation
rates.  Static finite-$G$ marginality and mean-field dynamical slowing are
therefore consistent at the endpoint.  Neither result establishes amorphous
localization, a force-bearing contact network, or Maxwell isostaticity.

The remaining question is consequently precise.  Why does loss of RY
stiffness at $\lambda_G^{T}=1$ coincide with exact terminal packing in one
dimension and occur near terminal disordered packing in two and three
dimensions?  A decisive test must compare the curvature zero and collective
slowing with independent real-space observables: suppression of $S(0)$,
near-contact scaling, contact number, and rigidity.  Agreement across
closures, lattice representations, dimensions, and compression protocols
would support a structural connection between the BCR endpoint and RCP;
failure would identify the numerical proximity as closure- or
representation-dependent.  The present result therefore sharpens, but does
not close, the passage from liquid-state correlations to terminal packing.

\end{document}